# Ion transport across solid-state ion channels perturbed by directed strain

A. Smolyanitsky*, A. Fang†, A.F. Kazakov, E. Paulechka

Applied Chemicals and Materials Division,
National Institute of Standards and Technology
Boulder, CO 80305

*To whom correspondence should be addressed: alex.smolyanitsky@nist.gov

## Abstract

We combine quantum-chemical calculations and molecular dynamics simulations to consider aqueous ion flow across non-axisymmetric nanopores in monolayer graphene and $MoS_2$. When the pore-containing membrane is subject to uniaxial tensile strains applied in various directions, the corresponding permeability exhibits considerable directional dependence. This anisotropy is shown to arise from directed perturbations of the local electrostatics by the corresponding pore deformation, as enabled by the pore edge geometries and atomic compositions. By considering nanopores with ionic permeability that depends on the strain direction, we present model systems that may yield a detailed understanding of the structure-function relationship in solid-state and biological ion channels. Specifically, the observed anisotropic effects potentially enable the use of permeation measurements across strained membranes to obtain directional profiles of ion-pore energetics as contributed by groups of atoms or even individual atoms at the pore edge. The resulting insight may facilitate the development of subnanoscale pores with novel functionalities arising from locally asymmetric pore edge features.

## Introduction

Control of solvated ion flow across nanoporous solid-state membranes is key to a wide range of nanofluidics applications, ranging from drug delivery [1] to energy storage [2]. In most cases, external gating of ionic permeation and selectivity is enabled by modifying the ion-pore interactions and/or the local thermodynamics. As a result, the main gating mechanisms include electrostatic gating [3-6], enzymatic-like processes [7], or changes in system temperature [8, 9]. In addition, 2D-material-based ion channels sensitively gated by tensile mechanical strain applied to the membrane were recently predicted [10, 11].

† Present address: BIOVIA, Dassault Systèmes, San Diego, CA, USA.

Strain-gated control of ion flow across crown-like pores in graphene [10, 12] and subnanoscale atomically symmetric multivacancies in molybdenum disulfide ($MoS_2$) [11] generally owes to the exceptionally high pore confinement experienced by the permeating ions. For ion-trapping pores (*e.g.*, graphene-embedded crown ethers), ion-pore and ion-solvent electrostatic interactions actively compete inside the pore and this competition is controlled by moderate pore dilation. As a result, order-of-magnitude permeability changes in response to few-percent strain are observed. For pores that do not trap ions, general ion-pore repulsion is reduced by pore dilation, also resulting in high mechanosensitivity [11].

In the case of isotropic strains applied to ideal crown pores featuring hexagonal edge symmetry, the edge atoms essentially form a circle around the ion located in the center of the pore. Consequently, pore dilation is assumed to be radial throughout the pore edge. Moreover, hexagonal symmetry yields nearly identical permeability responses to *uniaxial* tensile strains applied along different directions [10]. More generally, the assumption of a nearly circular pore edge underlies the well-accepted paradigm of ion-pore interactions, in which the pore is assumed to interact with permeating ions as a single isotropic entity. In the case of a non-axisymmetric pore edge lining, however, isotropy is not expected. The asymmetry within the pore edge can arise from different electrostatic charges carried by the edge atoms regardless of the pore shape, as well as from non-circular edge geometries (*e.g.*, triangular [13-16] or diamond-shaped [11] pores), as discussed in detail later. Given that in experiments, subnanoscale pores may not always have an axisymmetric structure, characterizing ion-pore interactions beyond the assumption of a circular edge may lead to a more focused understanding of structure-function relationships in subnanoscale pores.

Here, using quantum-chemical calculations and molecular dynamics (MD) simulations, we demonstrate examples of pores with marked anisotropy in their response to uniaxial strain, depending on its direction. We discuss the basic properties of the observed anisotropy, as determined by electrostatic and geometric effects. Finally, we demonstrate that permeation measurements across membranes subject to perturbative strains applied along various directions potentially enable obtaining directional profiles of the ion-pore interactions beyond the assumption of a uniform pore edge. In particular, for pores where short-range ion-pore interactions dominate, the resulting profiles are shown to contain features contributed by groups of atoms at the pore edge.

Ionic flow across various porous membranes was MD-simulated here, and one representative example is shown in Fig. 1. The system consists of a 5.6 nm × 6.4 nm monolayer graphene membrane featuring nine identical subnanoscale pores. The membrane is immersed in a room-temperature (T = 300 K) aqueous solution of 0.5M KCl in a 5-nm-tall periodic simulation cell. The ion flow is driven by a constant electric field $E_z$ = 0.05 V/nm, roughly corresponding to a transmembrane voltage of 0.05 V/nm × 5 nm = 0.25 V. The resulting currents are calculated from ion fluxes as described in our previous works [10, 11, 17]. The simulation setup uses the standard OPLS-AA framework [18, 19], identical to those in our previous works [10, 17]. Uniaxial in-plane tensile strains are applied to the membrane as described earlier [10] and in the SI. Strain magnitudes are varied up to 0.04, well within graphene's experimentally observed limits of elasticity [20].

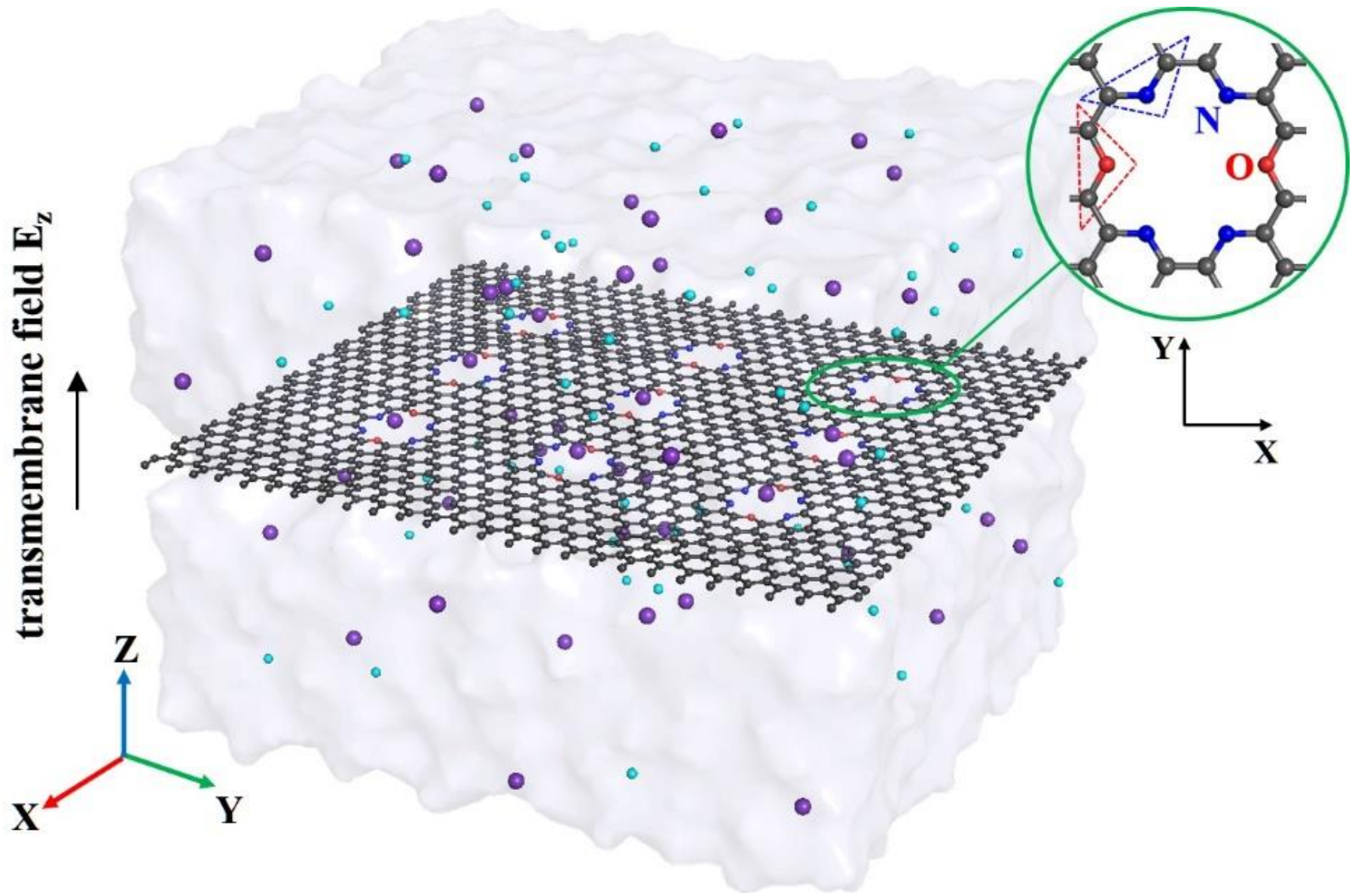

Figure 1. An example of simulated system: graphene membrane featuring nine pores in an aqueous ionic environment. Water and ions are shown as light-purple surface and bright spheres, respectively. The pore structure is shown in the upper right corner; the dashed triangles enclose the atoms forming pore edge dipoles.

As shown in the upper right corner of Fig. 1, the pore structure is similar to a graphene-embedded crown ether [17, 21] and is expected to trap solvated alkali metal ions. The important difference is that, instead of six edge oxygen atoms, only two edge atoms are oxygen atoms (red), while the rest are nitrogen atoms (blue). We refer to this pore as $N_4O_2$ in the text below. As discussed earlier [10, 17], ion-pore interactions are those between the ion and a total of six radially oriented dipoles formed by the edge atom (N or O) and the two neighboring carbons. Within the MD framework, the dipole, as enclosed in the dashed triangles in Fig. 1, is described by the atomic charge $Q_i$ of the edge atom ($i$ is N or O) and two $-Q_i/2$ charges of the two nearest carbons. If we posit that $Q_O$ and $Q_N$ are sufficiently different due to the differences in N-$C_2$ and O-$C_2$ bonding, there should arise a considerable anisotropy in the distortion of pore electrostatics, depending on the direction of externally applied uniaxial strain. On geometric grounds, for the pores in Fig. 1, this anisotropy should be maximal between strains applied along X and Y. Reasonable estimates of the atomic charges at the pore edge are key here. In addition to generally governing permeability [22] and mechanosensitivity [12], these charge values should directly control the degree of potential anisotropy. We therefore used the Gaussian 16 package [23] to perform quantum-chemical calculations and obtain the partial atomic charges according to the OPLS-AA-compatible CHELPG scheme [24]. The calculations were performed at the *HF/6-31+G(d)* theory level [25, 26]. The obtained charges (in the units of elementary charge) are $Q_O = -0.26$ and $Q_N = -0.64$, while each atom in the corresponding nearest-neighbor carbon pair carries charge of +0.13 and +0.32, respectively. Our main results were obtained with these charges. In addition, using the CP2K package [27], we performed DFT calculations to obtain the DDAP atomic charges [28]. These calculations, set up using the PBE exchange functional [29], Gaussian plane-wave pseudopotentials [30], and the DZVP basis set [31], yielded $Q_O = -0.23$ and $Q_N = -0.48$. The results obtained with these charges are provided in the section S1 of the Supporting Information (SI). We note that in our MD simulations the values of partial atomic charges (from both CHELPG and DDAP) are assumed to be constant with respect to the membrane strain. This assumption greatly simplifies simulation setup and enables a standard MD workflow. As estimated in section S1 of the SI for the DDAP charges, average atomic charge magnitudes in the DFT simulations varied by ~3.2% per 1% of isotropic strain. However, we expect that this level of charge variability would be unlikely to affect the qualitative nature of the results presented here [22].

**Results**

The ionic currents across graphene-embedded $N_4O_2$ pores described above were simulated for various strains, and the results are presented in Fig. 2. Indeed, the $N_4O_2$ structure exhibits different levels of mechanosensitivity, depending on the direction of strain, while biaxial isotropic strain expectedly causes the largest changes in ionic permeation. As described in our previous work [10], an estimate of the ion current as a function of strain ε is $I(\varepsilon) = I_0 e^{\mu\varepsilon}$, where $\mu = \frac{1}{k_B T}\left(\frac{dU}{d\varepsilon}\right)$ is dimensionless sensitivity to strain. Above, $U \approx U_{ion-pore} + U_{ion-water}$ approximates the total free energy of electrostatic interactions for an aqueous ion trapped in the pore; $k_B$ and $T$ are the Boltzmann constant and system temperature, respectively. In the differential limit of small strains, $\left(\frac{dU}{d\varepsilon}\right)$ is assumed to be constant. Fitting the data in Fig. 2 yields $\mu_{XX} = 45.80$, $\mu_{YY} = 68.57$, and $\mu_{XY} = 102.54$. The sensitivity to biaxial strain $\mu_{XY} \approx \mu_{XX} + \mu_{YY}$ is well-expected and consistent with earlier results [10], while the difference between $\mu_{YY}$ and $\mu_{XX}$ is of our current interest.

The curves fitted to XX and YY data in Fig. 2 yield an ion permeation anisotropy factor $\kappa_{YX} = I_{YY}/I_{XX}$ reaching ~3 at higher strains, consistent with $\kappa_{YX} \approx e^{(\mu_{YY}-\mu_{XX})\varepsilon} = 2.5$ at ε = 0.04. To describe the underlying mechanism we recall that, from the definition above, the Y-X permeation anisotropy can be expressed as

$$\kappa_{YX} \approx e^{\frac{\delta U_{YX}}{k_B T}}, \qquad (1)$$

where

$$\delta U_{YX} = \left(\delta U^{YY}_{ion-pore} - \delta U^{XX}_{ion-pore}\right) + \left(\delta U^{YY}_{ion-water} - \delta U^{XX}_{ion-water}\right) \qquad (2)$$

is the asymmetry in the corresponding free energy's response to the selected uniaxial strain directions. A naïve guess is that the ion-water asymmetry component in Eq. (2) is negligible. Given the data in Fig. 2, $\delta U_{YX}$ should not exceed $k_B T \approx 2.5$ kJ/mol within the presented strain range. However, the ion-pore component reaches $\left(\delta U^{YY}_{ion-pore} - \delta U^{XX}_{ion-pore}\right)$ ~ 10 kJ/mol at ε = 0.04 (see section S2 in the SI), corresponding to $\kappa_{YX} \approx 55$, far in excess of the data in Fig. 2. We therefore believe that $(\delta U^{YY}_{ion-water} - \delta U^{XX}_{ion-water})$ significantly opposes the "vacuum" ion-pore component. This hypothesis is supported by the MD-simulated ion-pore and ion-water energy asymmetries in Fig. 3. Because the ion-water energies have an average estimated uncertainty of $\sim 2k_BT$, an independent measure of ion-water interactions is necessary. It is presented in the inset of Fig. 3 in the form of ion-water first shell coordination numbers $N_c$, obtained as integrals of the corresponding radial distribution functions (RDFs) (see section S3 of the SI for details). As shown, $N_c^{YY} - N_c^{XX}$ increases with ε, confirming the ion-water energy trend.

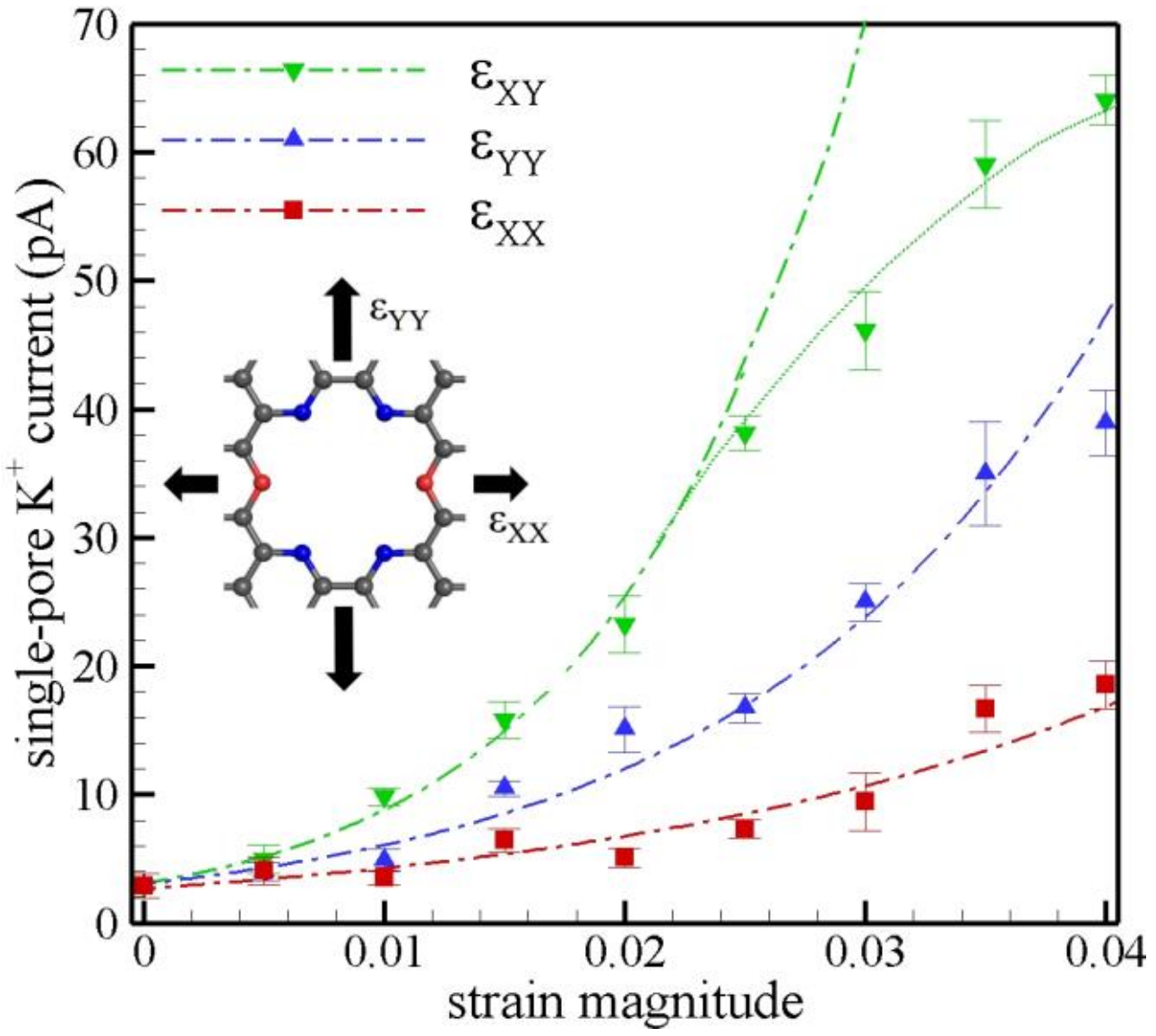

Figure 2. Single-pore $K^+$ currents as a function of membrane strains applied along the X, Y, and XY directions, where XY indicates isotropic ($\varepsilon_{XX} = \varepsilon_{YY}$) biaxial stretching. Each ion current value was obtained from the ionic flux simulated for 150 ns. No $Cl^-$ permeation was observed. The dash-dotted lines are exponential fits to simulated data. For XY-stretching, only the data points corresponding to $\varepsilon \leq 0.025$ were used for fitting due to saturation of the mechanosensitive effect for $\varepsilon > 0.025$ (dotted green curve).

Granted, it has been shown that the solvent moderates the overall level of mechanosensitivity in graphene-embedded crowns, whereby pore dilation leads to *reduction* in the ion-pore component and an *increase* in the ion-water component. The effect is essentially strain-dependent solvent screening in confinement [10]. Here, however, a direct argument based on the reduction of confinement is problematic, because on geometric grounds uniaxial membrane strains along X and Y direction dilate the pores nearly identically (see section S2 of the SI and Ref. [10]). The solvent's moderating effect on $\delta U$ has a more subtle explanation in this case. As estimated in the supplementary information of Ref. [10], $U_{ion-water}$ is an overall *increasing function of how much a pore-trapped ion fluctuates out of the pore plane*. Note that the RMSD of these fluctuations $\propto \sqrt{\frac{k_B T}{k_{ion-pore}}}$, where $k_{ion-pore}$ is the effective ion-pore force constant – an *increasing function of* the electrostatic potential well depth $|U_{ion-pore}|$. This RMSD is essentially a measure of how much the ion is thermally "peeking" into the solvent, which *increases with decreasing ion-pore interaction strength* [10]. Therefore, *any* reduction of the ion-pore interactions generally contributes toward increasing ion-water component. Consequently, anisotropy in the $U_{ion-pore}$ component *causes* a contribution toward opposing anisotropy in the $U_{ion-water}$ component. We also note that the anisotropy observed here is generally subject to thermal smearing and, given Eq. (1), is expected to decrease with increasing system temperature (see section S6 of the SI).

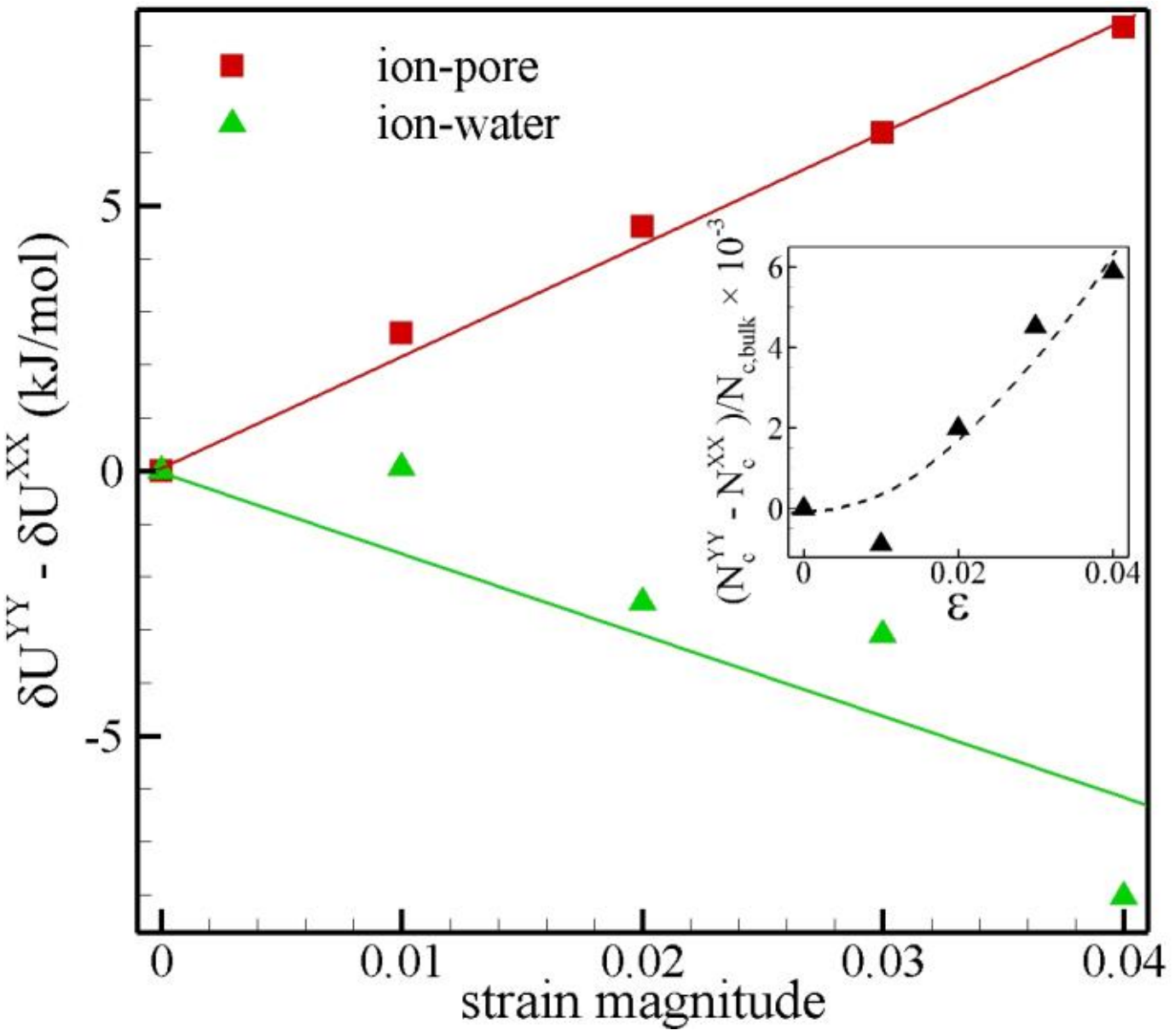

Figure 3. Simulated ion-pore and ion-water interaction energy anisotropies as functions of uniaxial strain. Each energy point was simulated for 50 ns. Solid lines are linear fits added as visual guides. All simulated energy values are negative (relative to vacuum), so the shown increasing and decreasing trends correspond to reduction and increase in interaction strengths, respectively. The inset shows anisotropy in the response of coordination numbers (integrated ion-water RDFs) to uniaxial strain.

So far, we have discussed uniaxial strains in the X and Y direction. In order to investigate $N_4O_2$ pores' response to uniaxial strain of a given magnitude applied at an arbitrary angle $\varphi$ (*e.g.*, relative to the X-direction), we have developed a procedure that deforms the membrane in an appropriate triclinic simulation box (see details in section S4 of the SI). Using this approach, we applied uniaxial strains in angle sweeps $0° \leq \varphi \leq 90°$ to cover the full angular range of the pore edge asymmetry. The results are shown in Fig. 4.

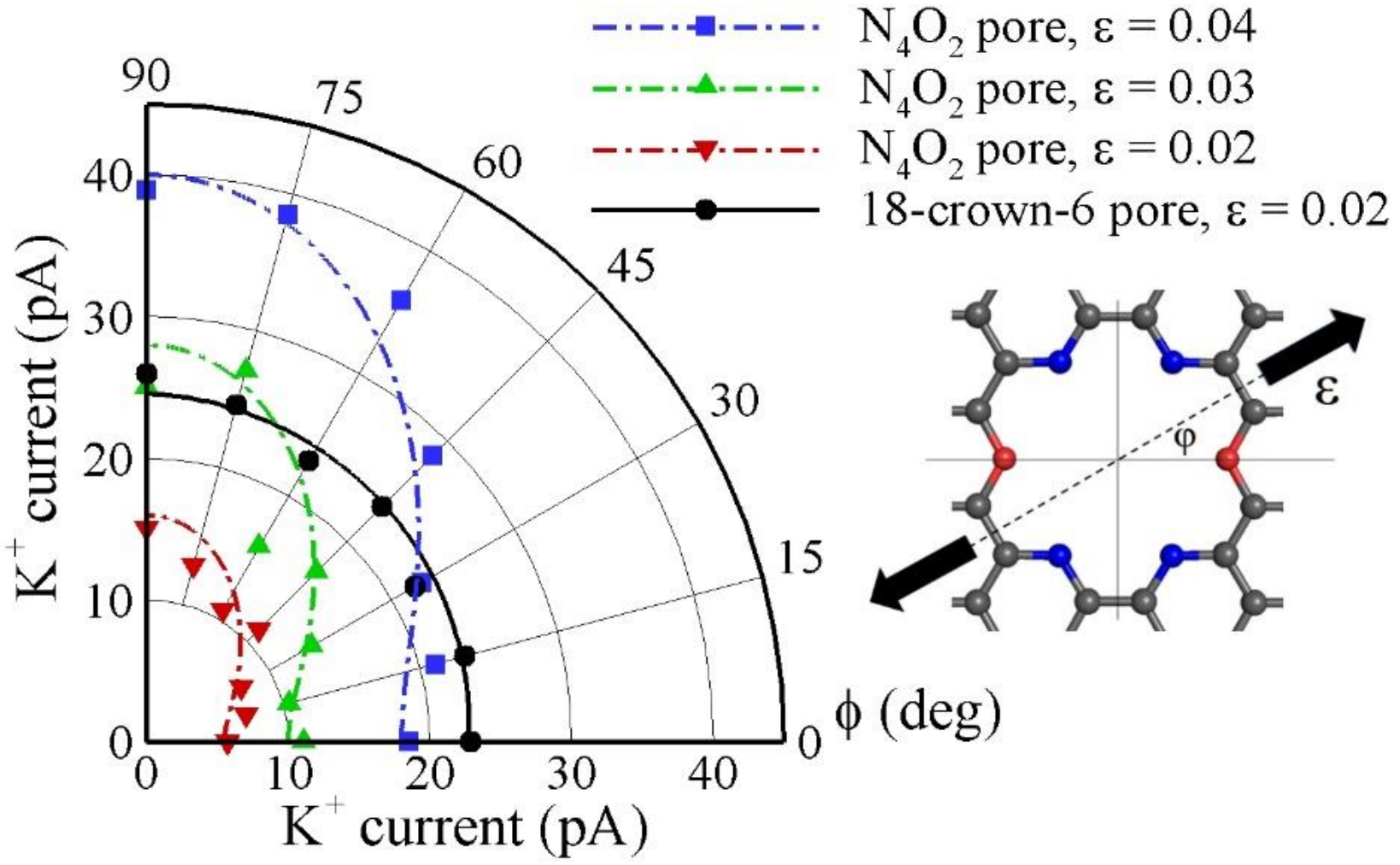

Figure 4. A family of polar curves describing $K^+$ permeation across graphene-embedded $N_4O_2$ pores as a function of uniaxial strain's direction, at various strain magnitudes. Each conduction point was simulated for 150 ns. In addition to the $N_4O_2$ pores, permeation *via* uniaxially strained 18-crown-6 pores is presented as an example of isotropic response (black circles and solid line). All continuous lines are $Aexp(Bcos\varphi)$-type fits to simulated data, added as visual guides. The corresponding data uncertainties are of the same order as the vertical bars shown in Fig. 2.

For all presented values of ε, the permeability of $N_4O_2$ pores increases continuously with increasing $\varphi$. In contrast, for the 18-crown-6 pore, the curve is nearly circular, expectedly corresponding to an isotropic case. Such a continuous response for the $N_4O_2$ pore is not surprising, given the gradually increasing electrostatic contribution from the displacements of nitrogen atoms carrying larger atomic charges, as $\varphi$ increases to 90°. Consider an ion trapped and traversing the pore near coordinates (0, 0) in the pore plane and surrounded by edge pore atoms indexed by $i$ and located at $\boldsymbol{r_i}$. The first-order electrostatic ion-pore component of $\delta U$ in response to an arbitrary uniaxial strain tensor $\boldsymbol{\varepsilon}$ is proportional to the projections of the atomic displacements upon the corresponding $\boldsymbol{r_i}$ as $\sum_i \frac{Q_i}{r_i^3}(\boldsymbol{\delta L_i}(\boldsymbol{\varepsilon}, \boldsymbol{r_i}) \cdot \boldsymbol{r_i})$. The sum is over all pore edge atoms. Above, $\boldsymbol{\delta L_i}(\boldsymbol{\varepsilon}, \boldsymbol{r_i})$ is the $i$-th atom's strain-induced displacement calculated from the corresponding *per-atom tensor*, specific to the mechanical properties of the pore edge in a given membrane material. For a pore with fully uniform in-plane response (of the isotropic host membrane, for instance), $\boldsymbol{\delta L_i}$ is calculated in the supplementary Eq. S11. Changes in the higher-order electrostatic contributions (*e.g.*, in the form of Lennard-Jones potential in MD simulations) have a similar dependence, except with higher powers of $r_i$ in the denominator. It is evident that, aside from the mechanical response of the material and the pore edge, the sources of anisotropy include both the charge values of the pore edge atoms $Q_i$ and the distances from the ion $r_i$. For ion-pore interactions dominated by Coulomb electrostatics, the directions of maximum response to uniaxial strain correspond to groups of atoms with the largest $Q_i$ magnitudes and the shortest $r_i$. One should not expect particularly sharp angular regions of minima or maxima in the response dominated by $1/r$ electrostatics, due to considerable contributions by *groups of edge atoms* and not individual atoms. As a result, the polar plots in Fig. 4 should be viewed as locally "smeared" energetic profiles of the pore edge.

It should now be clear that the directional response to uniaxial tensile strain arises from inhomogeneous pore edge charges and the pore geometry itself. Thus far, we have only considered directional anisotropy arising from a non-axisymmetric distribution of the relatively long-range first-order electrostatics along the pore edge. Non-circular pores in hexagonal boron nitride [13, 14] or triangular [15, 16] and diamond-shaped [11] pores in monolayer $MoS_2$ are marked by strong higher-order electrostatics and present another potentially interesting case. For example, let us consider the directional response of the diamond-shaped pores in $MoS_2$ [11]. As demonstrated earlier, these pores generally repel ions *and* feature significant atomic charge variation, depending on the edge atom location [11]. We performed an angular sweep of uniaxial strain ($\varepsilon = 0.04$) applied to a monolayer $MoS_2$ membrane featuring nine such pores. The simulation setup with 0.5M of aqueous NaCl was similar to that described above for graphene-embedded $N_4O_2$ pores and used a refined parameter set [32] utilized in our earlier work [11]. Pore geometry and the simulated angular range are shown in Fig. 5, along with the permeability data. Anisotropy of order ~3, similar to that in Figs 2 and 4, is observed. The angular range of permeation variability, however, is significantly narrower: the minima and maxima are separated by 30°, which is 1/3 of the corresponding range for $N_4O_2$ pores. This response cannot be explained based on the first-order electrostatics alone (in Fig. 5, atoms labeled **a** and **b** carry charges -0.022 and -0.33, respectively [11]). We therefore suggest that the short-range interactions, already known to underlie the repulsive nature of these pores [11], are mostly responsible for the relatively sharp angular features of the response. Consequently, for non-circular pores with a strong short-range component in the ion-pore interactions, the directional response to strain appears to yield ion-pore energy profiles with nearly atomistic detail.

Interestingly, such profiles may yield information about any parasitic pre-strain inevitably present in the pores. See section S7 of the SI for further details.

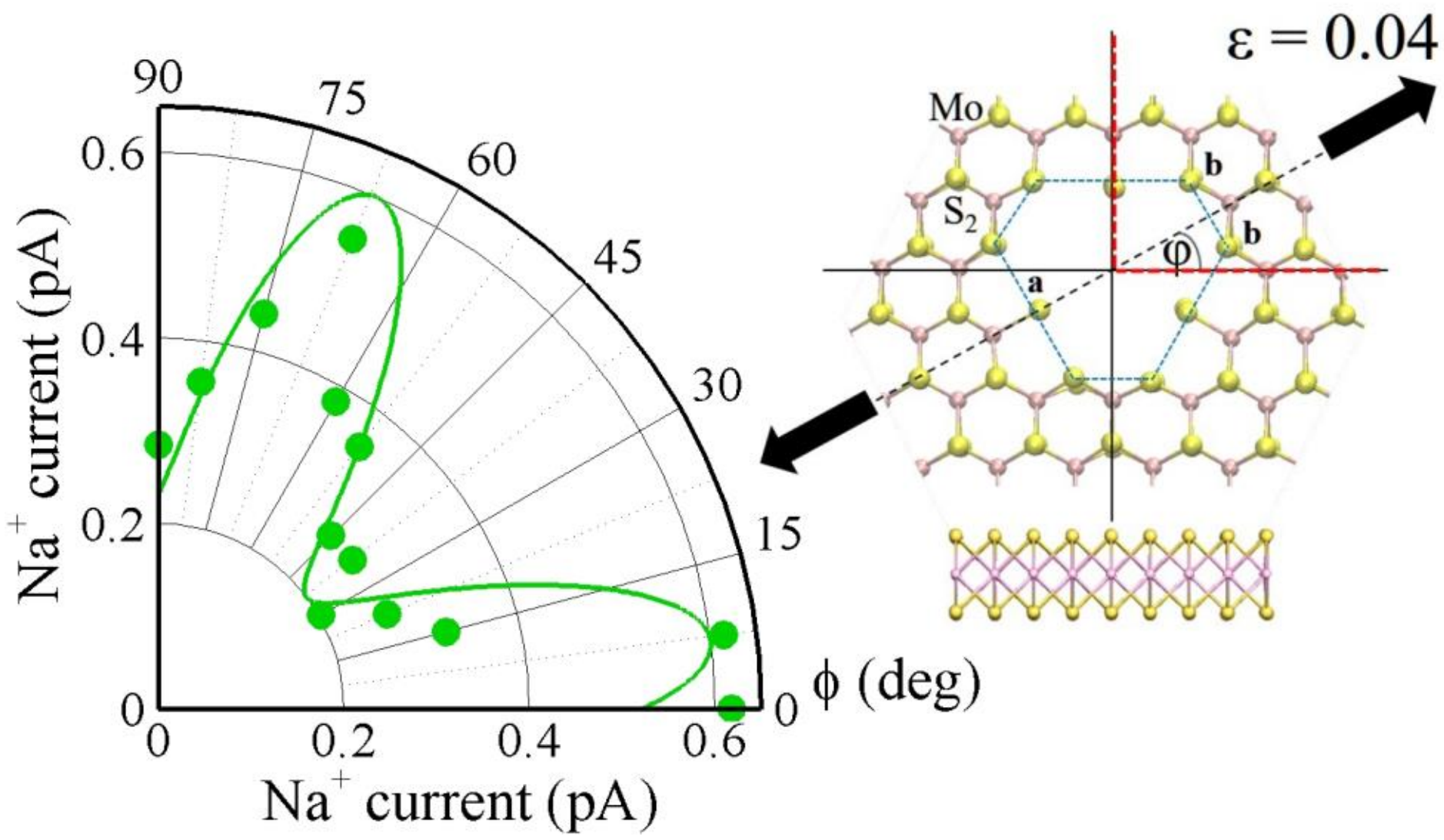

Figure 5. Ion-repulsive diamond-shaped pores in monolayer $MoS_2$ and the polar curve describing single-pore $Na^+$ permeation as a function of $\varphi$. The pore edge contour is provided by the dashed blue line. Permeation points were simulated at ε = 0.04 for 400 ns to 1600 ns per point and the average data uncertainty is 0.1 pA. The solid line is a visual guide of similar functional form as in Fig. 4. Dashed red lines outline the angular range of applied strains ($0° \leq \varphi \leq 90°$), which is roughly three times the observed response range.

**Conclusions**

We have shown that the ionic permeability of non-axisymmetric subnanoscale pores may have a sizeable dependence on the direction of uniaxial tensile strain applied to the membrane. The sources of anisotropy include inhomogeneity in the atomic charge distribution along the pore edge, as well as edge geometry. For the graphene-embedded hexagonal $N_4O_2$ pores that feature asymmetry in Coulomb interactions, the ion permeation is shown to change by a factor of ~3 within a 90-degree range of strain directions. For the diamond-shaped pores in monolayer $MoS_2$, where short-range repulsive interactions dominate, a similar degree of anisotropy is observed within a 30-degree range. As a result, nearly atomistic maps of the pore edge energetics may be obtained from macroscopic measurements of ionic permeability.

The obtained results suggest the possibility of "smart" membranes that significantly change their ionic permeability, depending on the direction of moderate tensile stretching. Because relatively minor energy changes ($\delta U \sim k_B T$) translate into sizeable current modifications *via* the corresponding Arrhenius exponent, our results may enable a path toward a detailed understanding of how ions interact with subnanoscale pores, including biological ion channels. In particular, the reported angular anisotropy suggests the possibility of probing energy contributions not only from the entire pore, but also *from groups of pore edge atoms, or even individual atoms*, depending on the pore geometry and the dominating type of interactions. In addition, exploring the moderating solvent effects on mechanosensitivity in general and in the context of anisotropy presented here may enable a better understanding of solvent screening in subnanoscale confinement. Finally, novel pore functionalities may be obtained from functionalization of portions of the pore lining, as well as from mere pore lining defects.

It is important to note that in membranes featuring pore arrays, observation of the phenomena predicted here requires not only control over the pore structure, but also identical/equivalent pore orientation relative to the host lattice. Therefore, without doubt, it will be challenging to experimentally verify our predictions, both in terms of membrane fabrication and precise application of strains. Nonetheless, recent advances in fabrication [15, 16] and strain actuation at the nanoscale [20, 33] suggest that rapid progress is being made toward the possibility of experimentally exploring strain-gated nanofluidic systems.

## Acknowledgments

The authors thank K. Kroenlein for devising and implementing data processing software. We are also grateful to C. Muzny and D. Luchinsky for enlightening discussions.
Part of this research was performed while A. Fang held a National Research Council (NRC) Postdoctoral Research Associateship at the National Institute of Standards and Technology (NIST). Authors gratefully acknowledge support from the Materials Genome Initiative.

## Disclaimer

Supplementary Information

for

**Ion transport across solid-state ion channels perturbed by directed strain**

A. Smolyanitsky, A. Fang, A.F. Kazakov, E. Paulechka

## S1. DDAP atomic charges: anisotropic permeation across $N_4O_2$ pores and charge value dependence on strain

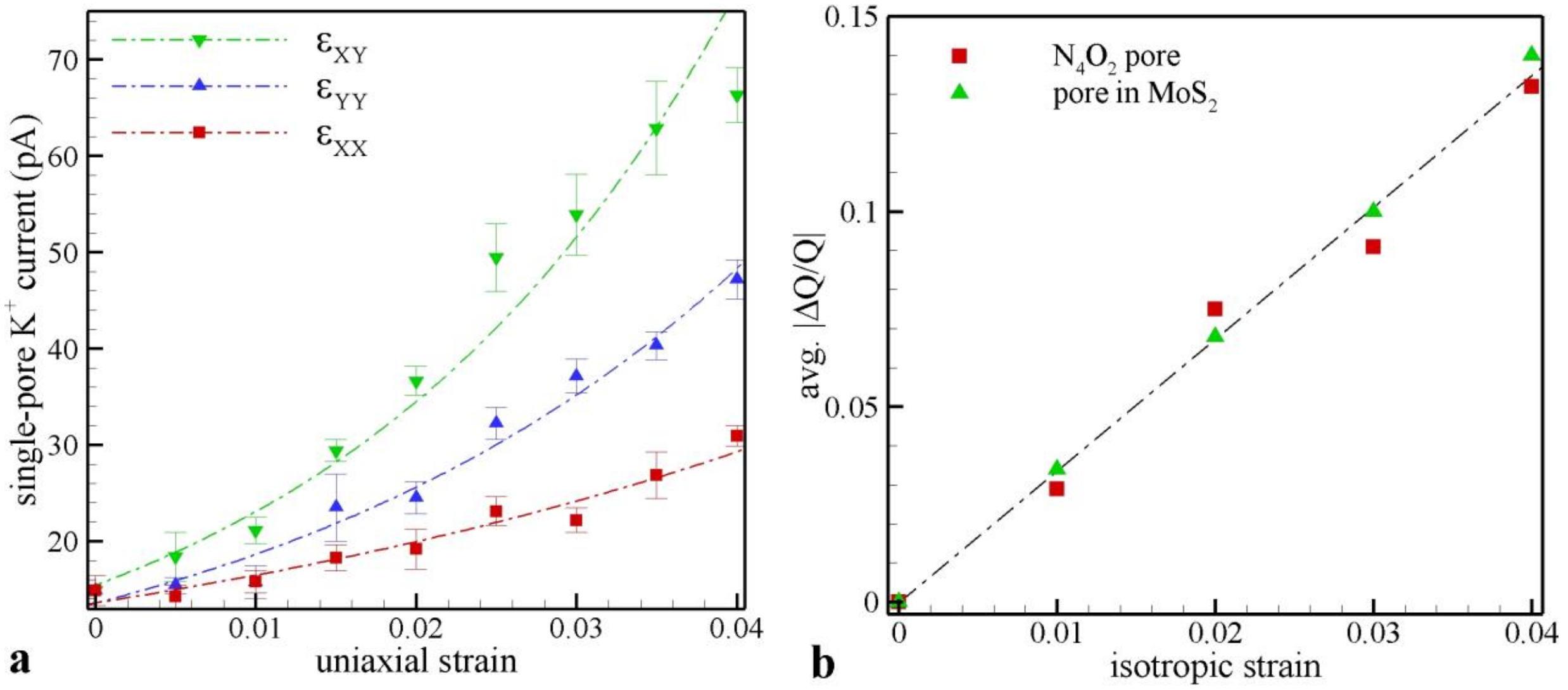

Figure S1. Analog of Fig. 2 in the main text, as obtained with DDAP charges ($Q_O$ = -0.23, $Q_N$ = -0.48) (a) and relative changes in DDAP atomic charges (averaged throughout the corresponding set of atoms lining each pore) as functions of isotropic strain (b). In (a), the strain sensitivities are $\mu_{XX} = 19.14$, $\mu_{YY} = 31.74$, and $\mu_{XY} = 48.03$ ($\mu_{XY}$ is obtained from exponential fitting to the data corresponding to $\varepsilon \leq 0.025$). In (b), atomic charges tend to *decrease* (in absolute value) with increasing strain.

## S2. Analytical estimates of permeation anisotropy

The level of anisotropy in ionic permeability gated by uniaxial strains can be roughly estimated analytically. In the simplest case of X and Y strain directions, consider the structure in Fig. 1. The free energy for a pore-trapped solvated ion interacting with the hydrated pore is approximated as [1]:

$$U = U_{ion-pore} + U_{ion-water}, \quad \text{(S1)}$$

where $U_{ion-pore} \propto \sum \left( \frac{Q_i}{r_0} - 2\frac{Q_i/2}{r_1} \right)$ is the sum of vacuum interactions with six edge dipoles (indexed by *i*) shown in Fig. 1. Here, each dipole consists of the negative atomic charge $-Q_i$ of the nitrogen or oxygen atom at the inner edge of the pore and a corresponding pair of carbons with charges $\frac{+Q_i}{2}$ at the outer edge. The following estimate does not include anisotropy in the changes $U_{ion-water}$ between Y and X directions and is limited to estimating the $\delta U_{ion-pore}^{YY} - \delta U_{ion-pore}^{XX}$ in the differential limit of pore geometry distortions, assuming that all atomic displacements due to strain correspond to bulk continuum values and ignoring the mechanical properties of the pore region. Also, only first-order electrostatic energy terms are included in the calculations. Just like in our MD simulations, we assume that the atomic charge values remain constant with respect to bond stretching considered here. Given these assumptions, the following is a simple exposition of the main source of anisotropy and not a quantitative estimate.

For simplicity, absolute values of atomic charges are used below.

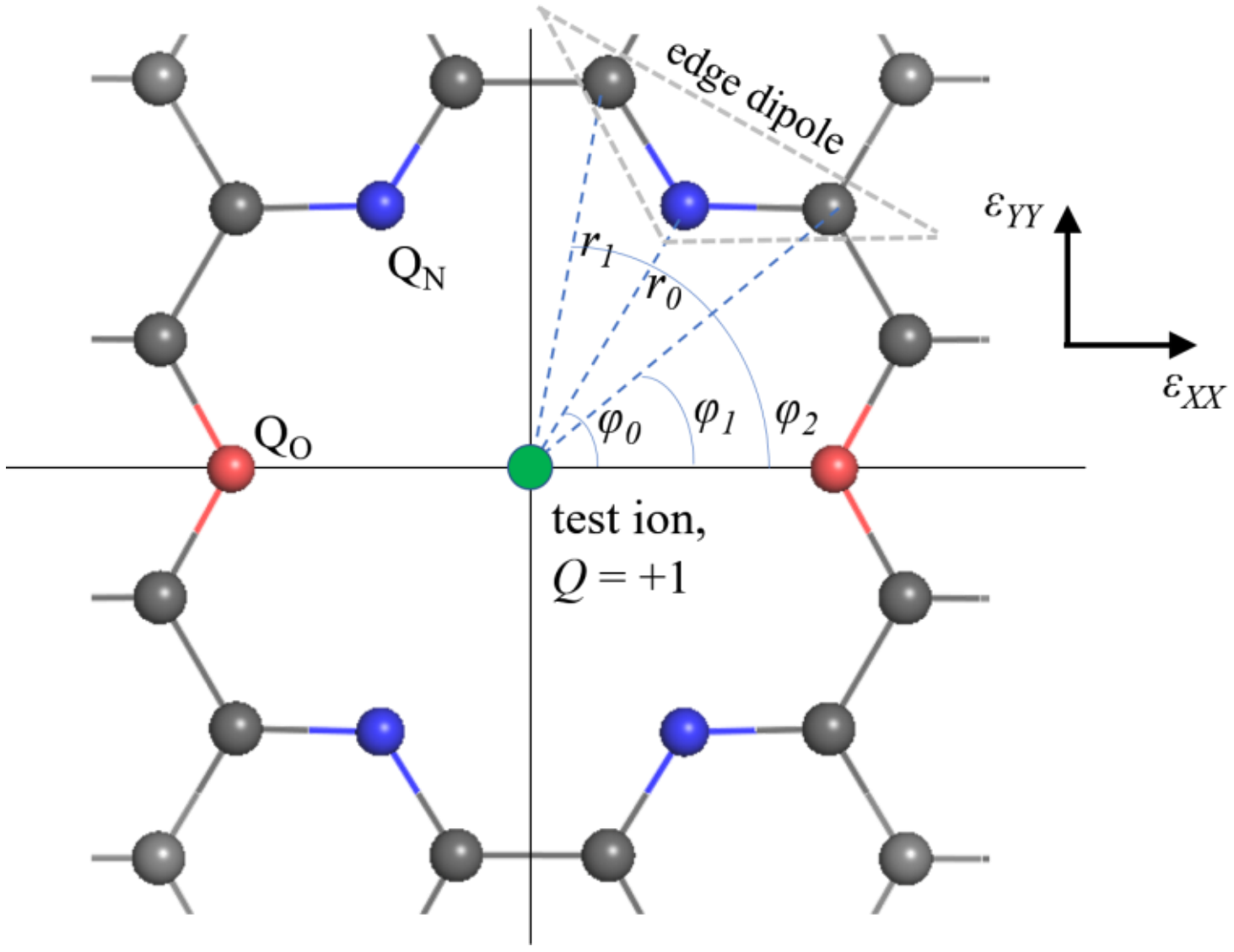

Figure S2. A positively charged test ion in an $N_4O_2$ crown-like pore subject to uniaxial strain along X or Y direction.

In general, for an arbitrary uniaxial strain direction, the change in $U_{ion-pore}$ is effectively the corresponding sum of tensor elements describing the changes in $r_0$ and $r_1$ for all dipoles interacting with the test ion in Fig. S2. With pore geometry in Fig. 1, each element can be estimated directly. For any membrane atom in the absence of strain interacting with the test ion at a distance $r$ with the corresponding radius-vector $\boldsymbol{r}$ forming an angle $\varphi$ with the X-direction, the radii perturbed by small uniaxial strains $\varepsilon_{XX}$ and $\varepsilon_{YY}$ are, respectively:

$$r_{XX} \approx r(1 + \varepsilon_{XX}\, cos^2\varphi),\ r_{YY} \approx r(1 + \varepsilon_{YY}\, sin^2\varphi). \quad \text{(S2)}$$

For uniaxial $\varepsilon_{YY}$, all inner edge N atoms are vertically displaced, along with the corresponding outer-edge carbons, above and below Y = 0. The two O atoms remain unperturbed, while the corresponding carbons are displaced. The resulting change in energy is

$$\delta U_{ion-pore}^{YY} \propto \frac{3Q_N}{r_0}\varepsilon_{YY} - \frac{39}{14}\frac{Q_N}{r_1}\varepsilon_{YY} - \frac{3}{14}\frac{Q_O}{r_1}\varepsilon_{YY}. \quad \text{(S3)}$$

For uniaxial $\varepsilon_{XX}$, all inner and outer atoms are displaced horizontally to the left and right of X = 0, so that:

$$\delta U_{ion-pore}^{XX} \propto \frac{Q_N}{r_0}\varepsilon_{XX} - \frac{17}{14}\frac{Q_N}{r_1}\varepsilon_{XX} + \frac{2Q_O}{r_0}\varepsilon_{XX} - \frac{25}{14}\frac{Q_O}{r_1}\varepsilon_{XX}. \quad \text{(S4)}$$

As a sanity check, adding Eqs. (S3) and (S4), we obtain the correct energy change in response to biaxial strain. Assuming equal uniaxial strain magnitudes $\varepsilon_{YY} = \varepsilon_{XX} = \varepsilon$, we obtain:

$$\delta U_{ion-pore}^{YY} - \delta U_{ion-pore}^{XX} \propto \frac{2(Q_N - Q_O)}{r_0}\varepsilon - \frac{11}{7}\frac{(Q_N - Q_O)}{r_1}\varepsilon =$$

$$= 2(Q_N - Q_O)\left(\frac{1}{r_0} - \frac{1}{r_1}\right)\varepsilon + \frac{3}{7}\frac{(Q_N - Q_O)}{r_1}\varepsilon. \quad \text{(S5)}$$

Eq. (S5) without the rightmost term sets the following lower limit on the anisotropy in energy:

$$\delta U_{ion-pore}^{YY} - \delta U_{ion-pore}^{XX} \propto 2(Q_N - Q_O)\left(\frac{1}{r_0} - \frac{1}{r_1}\right)\varepsilon. \quad \text{(S6)}$$

Note that the right side of Eq. (S6) is conveniently equal to 1/3 of the change in total energy of ion-pore interaction in a *fully axisymmetric hexagonal pore lined with atoms carrying charge* $(Q_N - Q_O)$, subject to isotropic biaxial strain $\varepsilon$. This finding enables a rough numerical estimate.

To proceed from proportionality to functional dependence, we use the results in Ref. [1] to estimate the per-atom strain susceptibility $\mu_a \propto \left(\frac{1}{r_0} - \frac{1}{r_1}\right)$. From Fig. 4a therein, 2% of isotropic strain applied to a fully symmetric hexagonal pore lined with six oxygen atoms (crown $|Q_O|$ = 0.4) results in ~6$k_bT$ = 15 kJ/mol change in the ion-pore electrostatic energy. Thus, *per unit of strain*, *per unit of charge of the inner edge atom*, the susceptibility is $\mu_a$ = 750. The ion-pore energy anisotropy is therefore *at least*

$$\delta U_{ion-pore}^{YY} - \delta U_{ion-pore}^{XX} = \frac{\mu_a k_b T}{3}(Q_N - Q_O)\varepsilon. \quad \text{(S7)}$$

Finally, we correct for the Poisson effect and introduce Poisson's ratio υ = 0.19 [2], corresponding to bulk graphene with ripples significantly suppressed by water. The reader is encouraged to confirm that after simple manipulations Eq. (S7) becomes

$$\delta U_{ion-pore}^{YY} - \delta U_{ion-pore}^{XX} = \frac{\mu_a k_b T}{3}(1 - \upsilon)(Q_N - Q_O)\varepsilon. \quad \text{(S7a)}$$

A comparison between MD-simulated data and Eq. (S7a) is shown in Fig. S3.

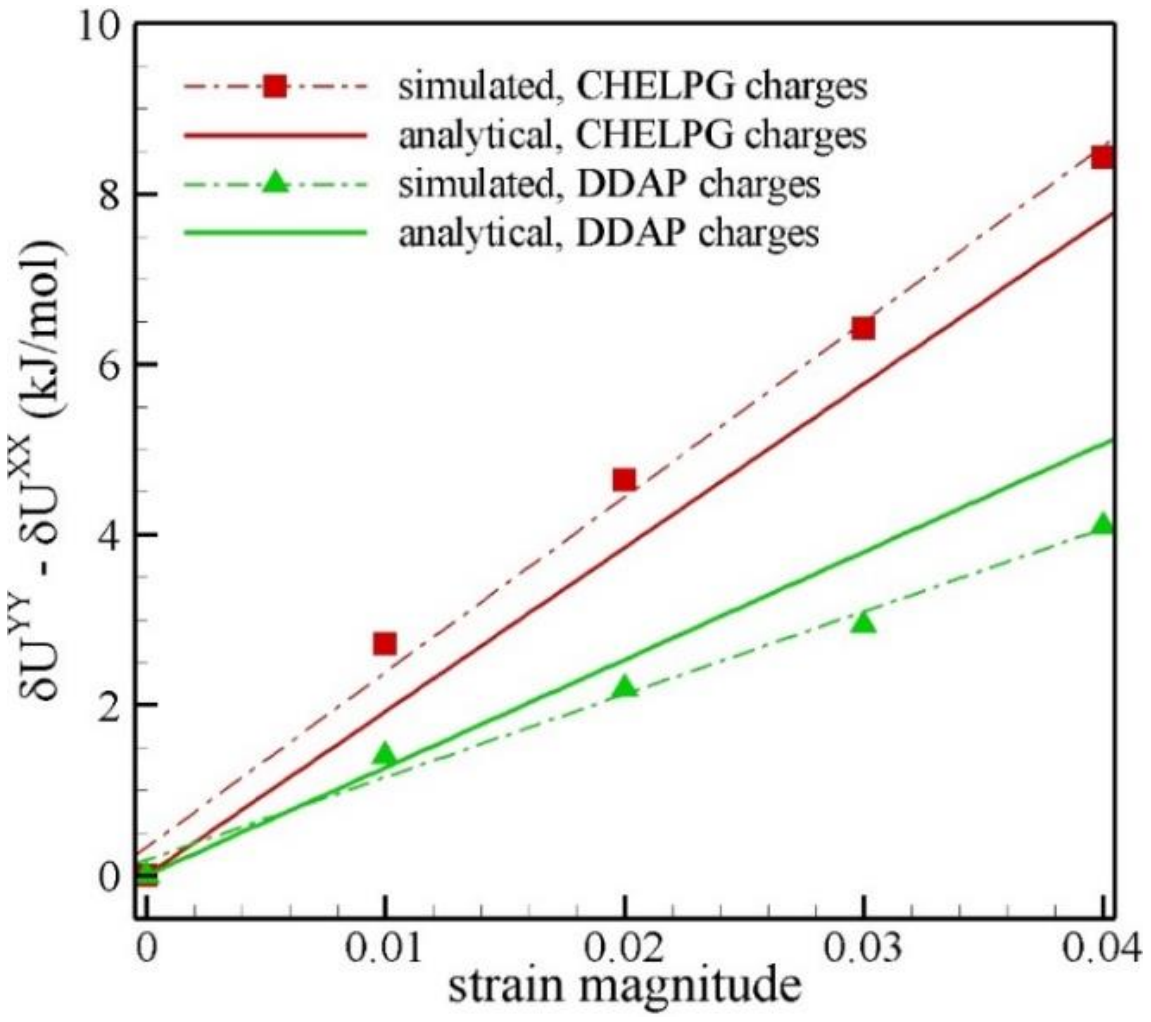

Figure S3. Simulated $\left(\delta U_{ion-pore}^{YY} - \delta U_{ion-pore}^{XX}\right)$ alongside the estimates by Eq. (S7a). The data is presented for CHELPG and DDAP atomic charge sets.

## S3. Ion-water coordination

The coordination numbers are calculated as follows: $N_c = 4\pi \int_0^{r_1} g(r) r^2 dr$, where $r_1 = 0.36$ nm approximately corresponds to the first hydration shell. Here, $g(r)$ is the ion-water-oxygen radial distribution function (RDF). Shown in Fig. S4 are representative examples of MD-simulated RDF curves for a $K^+$ ion in bulk water and the same ion trapped in the unstrained $N_4O_2$ pore, as well as this pore subject to $\varepsilon_{YY} = \varepsilon_{XX}$= 0.04 in the presence of water.

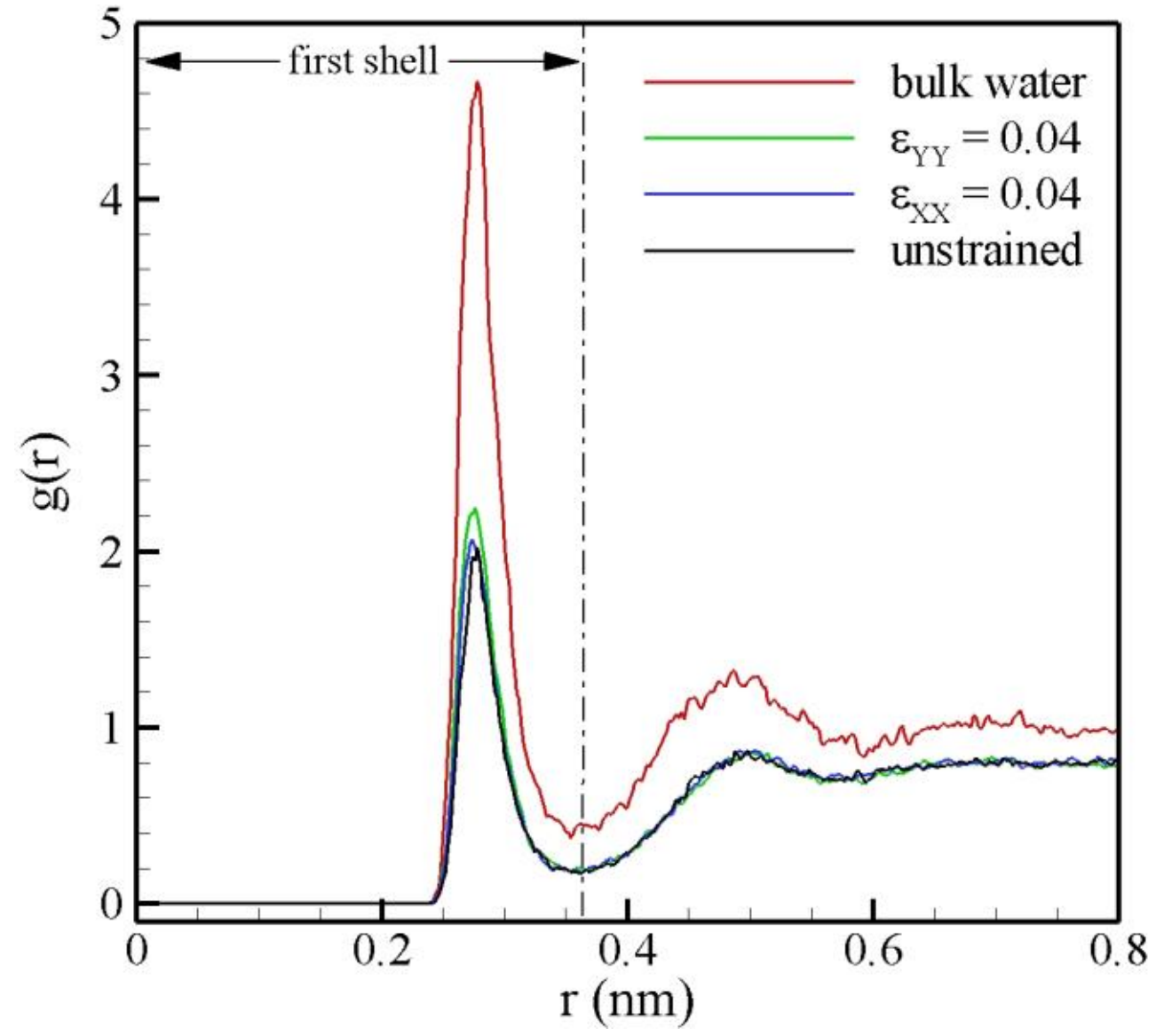

Figure S4. Examples of normalized ion-water-oxygen RDFs for $K^+$ ions in bulk water and in strained and unstrained $N_4O_2$ pores (each RDF is calculated from 2500 timeframes in a 50-ns-long simulation). As calculated, $N_{c,bulk}$= 6.95.

## S4. Uniaxial strain applied along an arbitrary direction using a triclinic simulation cell

Here, a uniaxial strain of magnitude $\varepsilon$ is directed at an angle $\varphi$ with the X-axis. In the following, we find the parameters that define the triclinic cell of the strained system and the transformed initial coordinates of the atoms in the triclinic system (see Fig. S5). The following should be applicable to any simulation cell setup for the Gromacs simulation package.

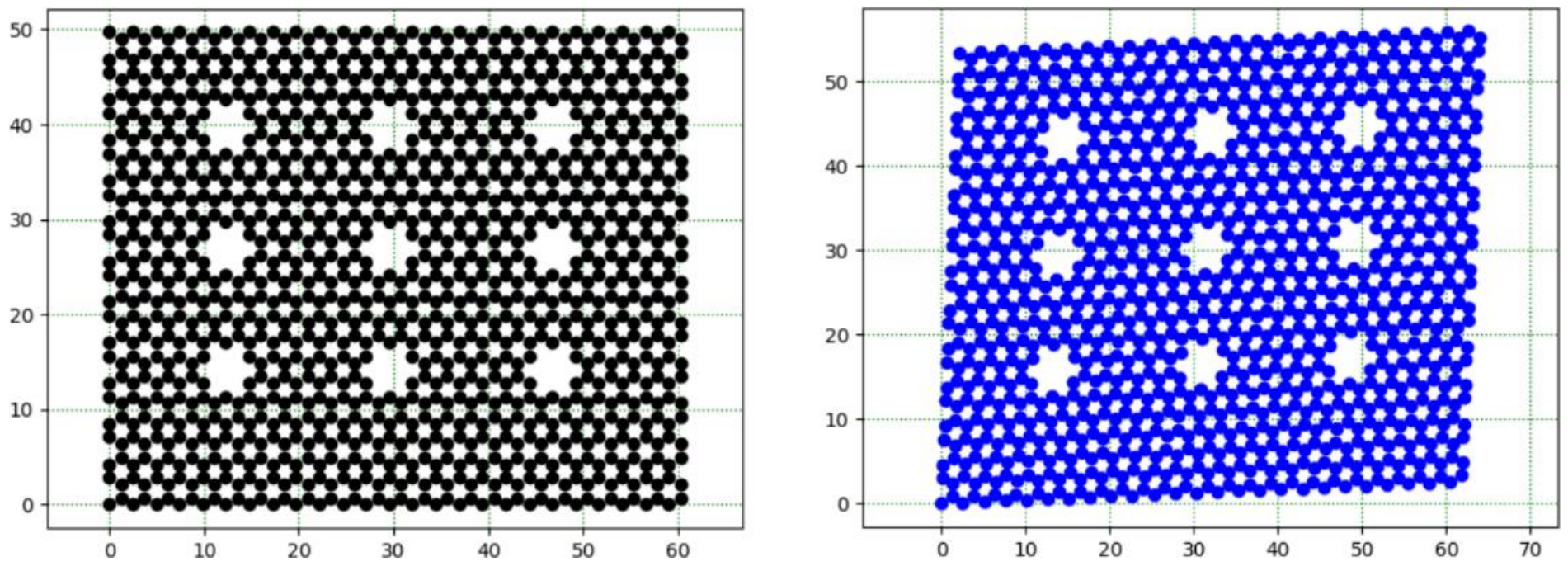

Figure S5. Example unstrained and strained triclinic system. The dimensions are in angstroms.

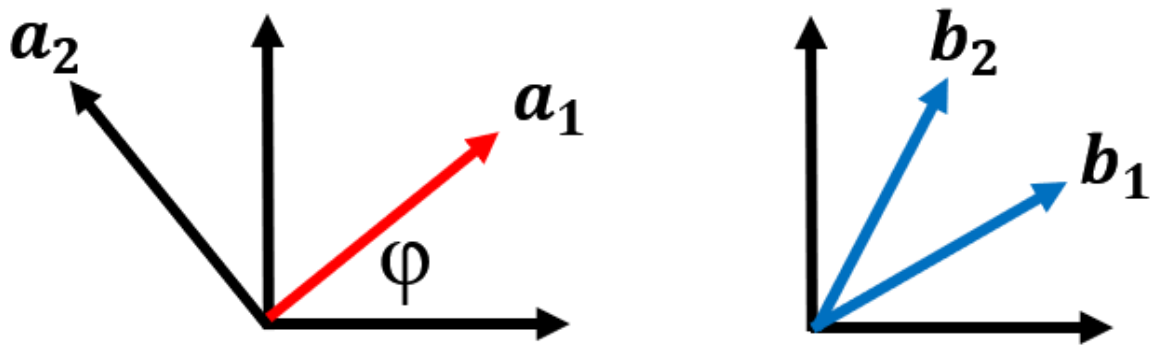

Figure S6. Vector definitions; strain is applied along $\boldsymbol{a_1}$.

The strain tensor converted from the basis oriented along the direction of uniaxial strain to the original coordinate axes is:

$$\begin{bmatrix} \varepsilon_{XX} & \varepsilon_{XY} \\ \varepsilon_{XY} & \varepsilon_{YY} \end{bmatrix}_O = \begin{bmatrix} cos\varphi & sin\varphi \\ -sin\varphi & cos\varphi \end{bmatrix} \begin{bmatrix} \varepsilon & 0 \\ 0 & 0 \end{bmatrix}_A \begin{bmatrix} cos\varphi & -sin\varphi \\ sin\varphi & cos\varphi \end{bmatrix} = \quad \text{(S9)}$$

$$= \begin{bmatrix} \varepsilon \, cos^2\varphi & \varepsilon \, sin\varphi \, cos\varphi \\ \varepsilon \, sin\varphi \, cos\varphi & \varepsilon \, sin^2\varphi \end{bmatrix}. \quad \text{(S10)}$$

The coordinate transformation is then:

$$\begin{bmatrix} x \\ y \end{bmatrix} = \begin{bmatrix} \varepsilon \, cos^2\varphi & \varepsilon \, sin\varphi \, cos\varphi \\ \varepsilon \, sin\varphi \, cos\varphi & \varepsilon \, sin^2\varphi \end{bmatrix} \begin{bmatrix} x_0 \\ y_0 \end{bmatrix} + \begin{bmatrix} x_0 \\ y_0 \end{bmatrix}. \quad \text{(S11)}$$

Thus, the transformed unit vectors pointing along the triclinic cell edges can be found by setting (1, 0) and (0, 1) in Eq. (S11):

$$\overrightarrow{b_1} = \frac{1}{\delta_X} \begin{bmatrix} 1 + \varepsilon \, cos^2\varphi \\ \varepsilon \, sin\varphi \, cos\varphi \end{bmatrix}, \quad \text{(S12)}$$

$$\overrightarrow{b_2} = \frac{1}{\delta_Y} \begin{bmatrix} \varepsilon \, sin\varphi \, cos\varphi \\ 1 + \varepsilon \, sin^2\varphi \end{bmatrix}, \quad \text{(S13)}$$

where

$$\delta_X = \sqrt{(1 + \varepsilon \, cos^2\varphi)^2 + \varepsilon^2 sin^2\varphi cos^2\varphi}, \quad \text{(S14)}$$

$$\delta_Y = \sqrt{(1 + \varepsilon \, sin^2\varphi)^2 + \varepsilon^2 sin^2\varphi cos^2\varphi}. \quad \text{(S15)}$$

For a simulation box of original in-plane dimensions $L_X$ and $L_Y$, the new dimensions are $L_X\delta_X$ and $L_Y\delta_Y$. The angle between unit vectors is $\alpha = cos^{-1}(\overrightarrow{b_1} \cdot \overrightarrow{b_2})$. The transformed coordinates expressed in the triclinic system are $x_{tri} = \delta_X x$ and $y_{tri} = \delta_Y y$, and the new position vectors are $\vec{r} = x_{tri}\overrightarrow{b_1} + y_{tri}\overrightarrow{b_2}$ when expressed in the original coordinate system.

## S5. Directional response for diamond-shaped and triangular pores

To better illustrate the symmetry observed in the response of permeability to uniaxial strains, an example of the full set of equivalent strain directions is shown in Fig. S7. In addition, we performed a directional strain sweep applied to nitrogen-terminated triangular pores in monolayer hexagonal boron nitride (hBN) [3, 4]. Note that all atomic charges were set to their bulk values according to the recently reported parameterization [5] and thus the anisotropy here arises only from the pore geometry. The results are shown in Fig. S8; the observed anisotropy, although modest, exhibits symmetry similar to that observed for the diamond-shaped pores in $MoS_2$.

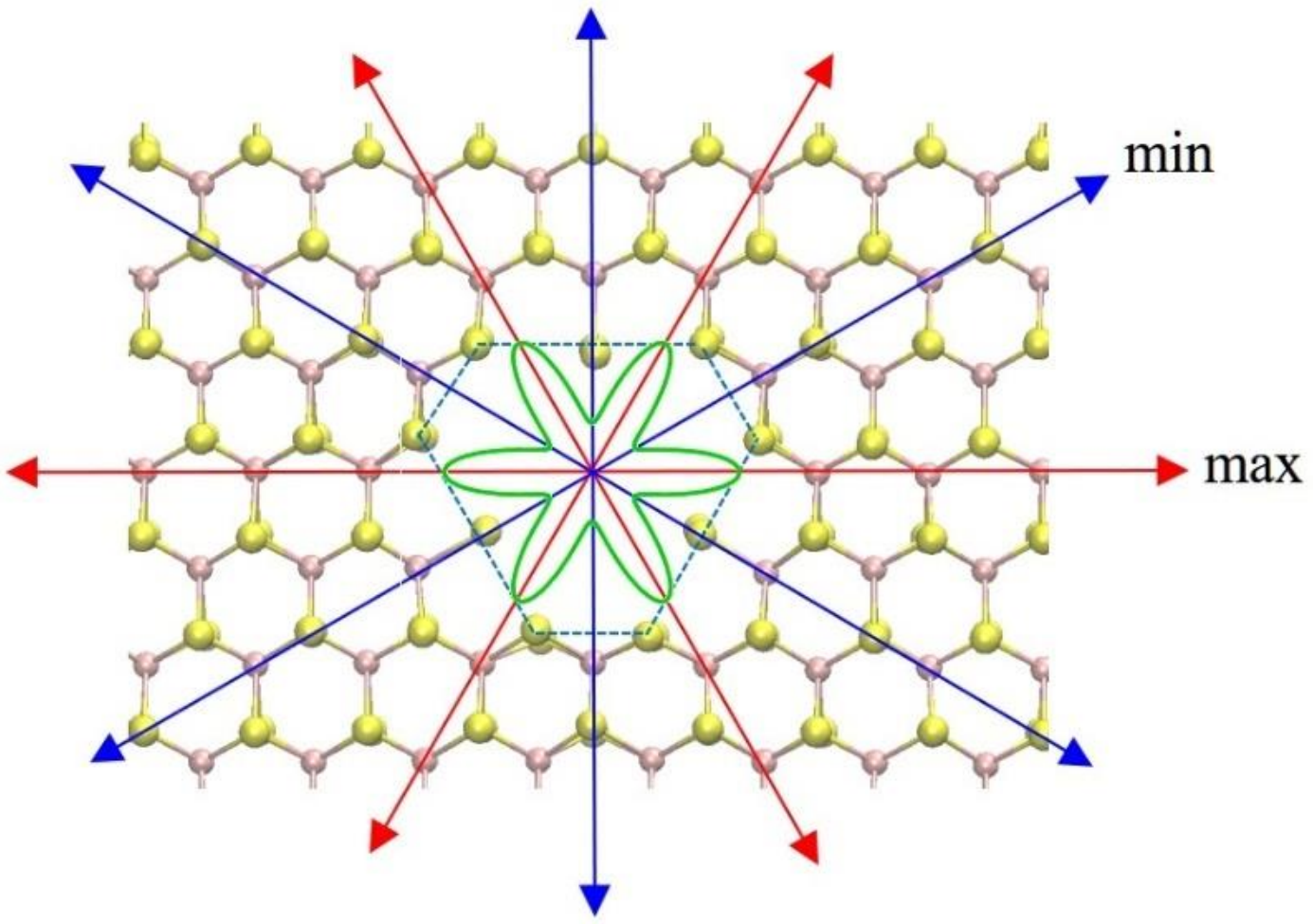

Figure S7. Diamond-shaped pore in $MoS_2$ and a complete set of equivalent directions corresponding to the angular response shown in Fig. 5 of the main text. The green "flower" corresponds to the sinusoidal data fit in Fig. 5. Note that for clarity the ~7-degree tilt in the response observed in Fig. 5 is omitted here. Blue and red arrows correspond to uniaxial strain directions with the minimal and maximal ionic currents, respectively.

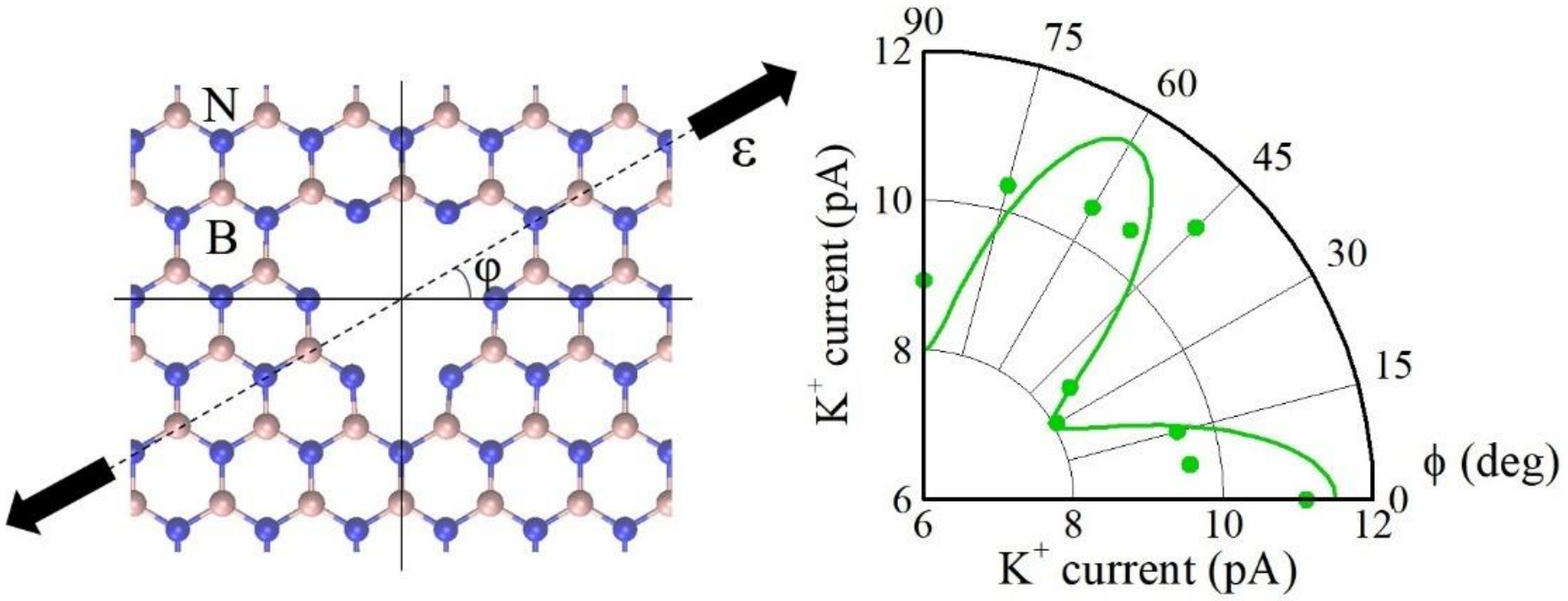

Figure S8. Single-pore $K^+$ currents across N-terminated triangular pores in monolayer hBN, as obtained from 0.5M aqueous KCl. Each permeation point was simulated for 400 ns. The directional strain sweeps were performed in the range $0° \leq \varphi \leq 90°$, as described in the main text. The average data uncertainty is 2 pA.

To demonstrate detection of parasitic pre-strain present in the membrane, we performed simulations similar to those presented in the main Fig. 5, except using a membrane, in which the pores are densely spaced and there is parasitic pre-strain ($\varepsilon \approx 0.01$) along the X-direction. The results of the angular sweep in Fig. S9, aside from revealing larger ionic currents than those in Fig. 5, contain a significant symmetry distortion in the *Y-direction* in the corresponding permeability response.

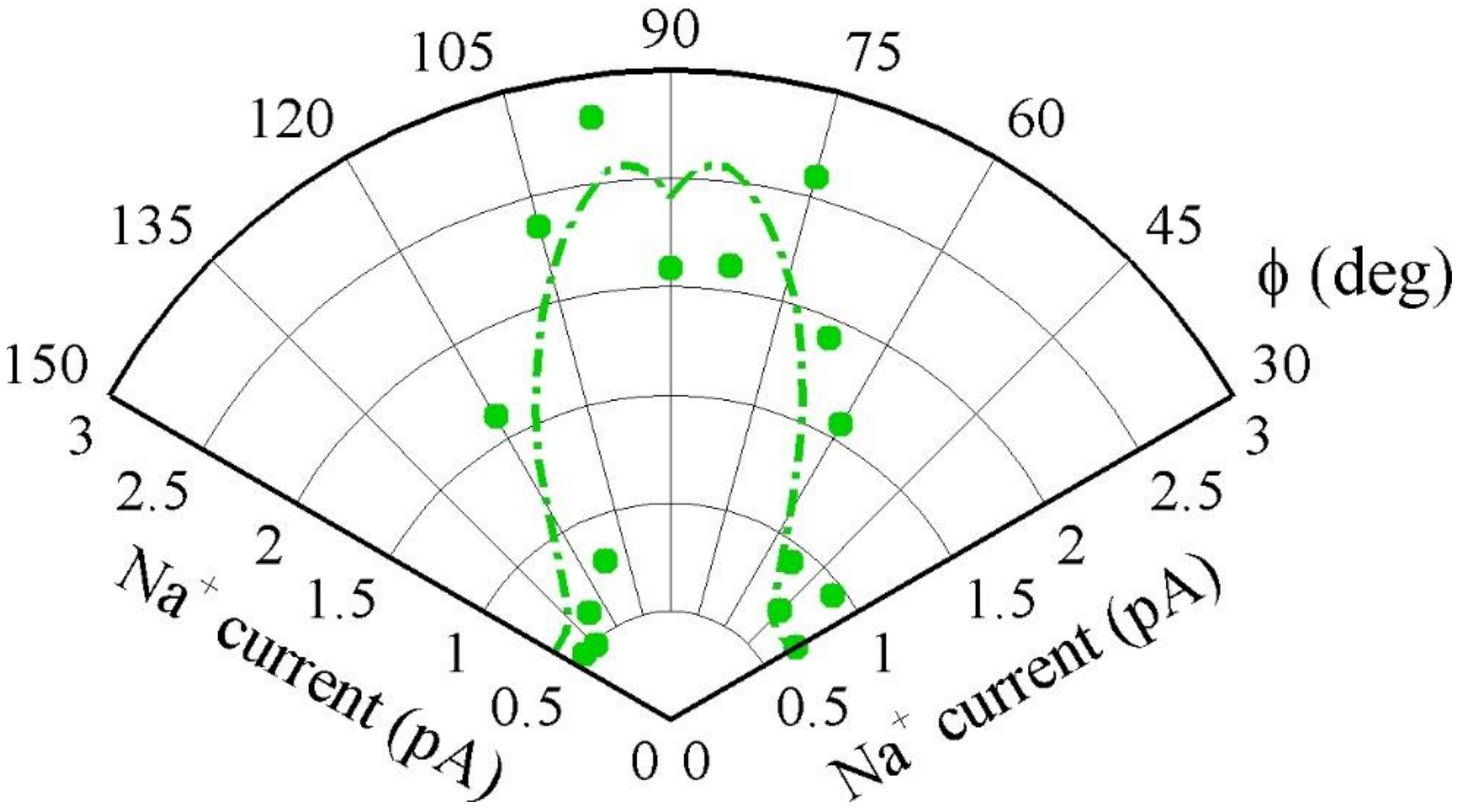

Figure S9. Response of $Na^+$ permeability to the angular sweep of uniaxial tensile strain ($\varepsilon = 0.04$), as applied to pre-strained pores. The angular range is $30° \leq \varphi \leq 150°$, roughly corresponding to twice the response period of the pores in Fig. 5 and Fig. S7. Each permeation point was simulated for 400 ns. The average data uncertainty is 0.2 pA.

**S6. Temperature dependence of permeation anisotropy in $N_4O_2$ pores**

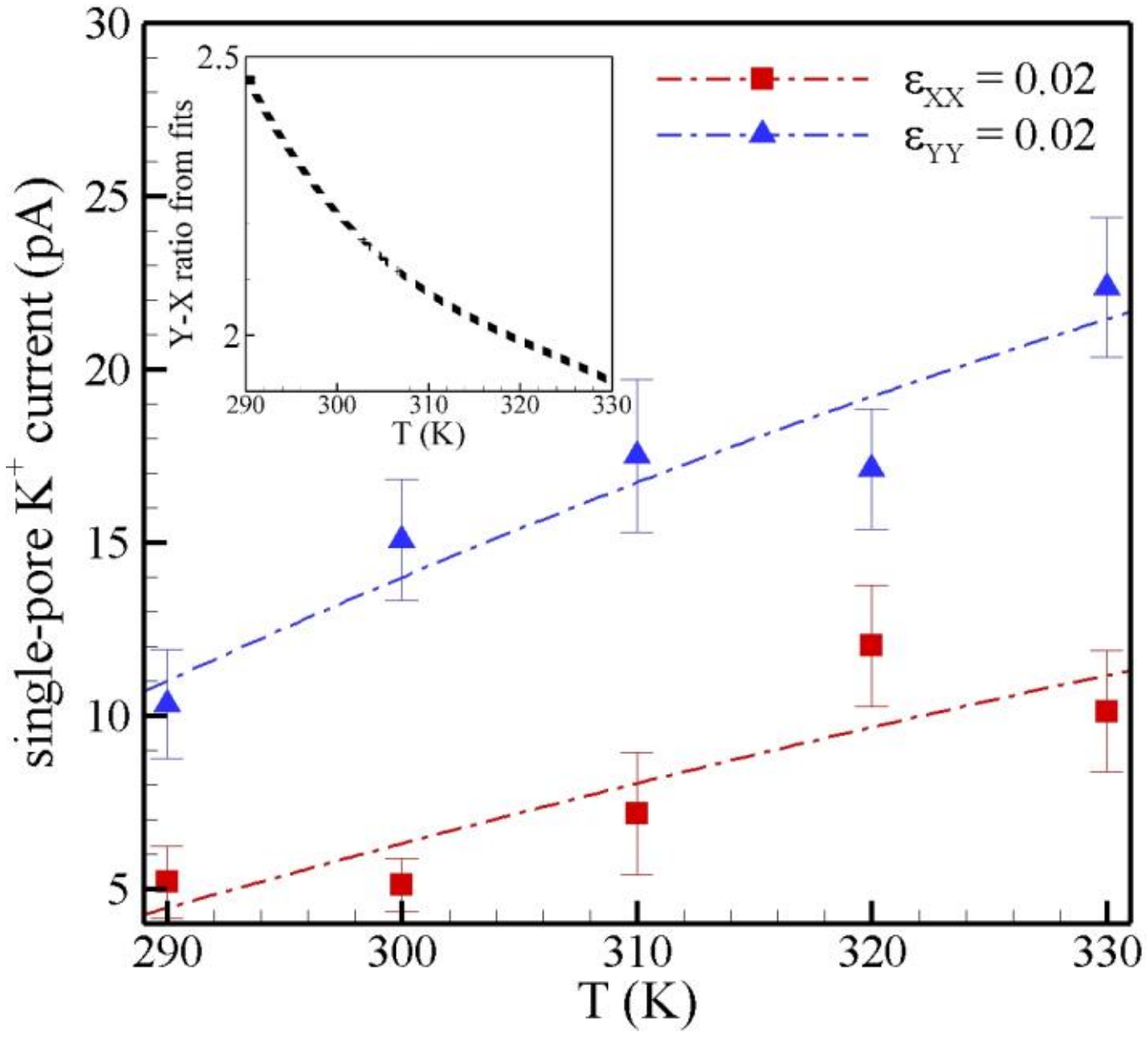

Figure S10. Single-pore $K^+$ currents across $N_4O_2$ pores subject to uniaxial strains $\varepsilon_{XX} = 0.02$ and $\varepsilon_{YY} = 0.02$ at various temperatures. The dash-dotted lines are quadratic fits to the ionic current data, used to obtain the trend in the resulting anisotropy ratio $\kappa_{YX} = I_{YY}/I_{XX}$ shown in the inset.